\documentclass[%
reprint, amsmath,amssymb,%twocolumn,
pra,
]{revtex4-1}

\usepackage{graphicx}
\usepackage{epsfig}
\usepackage{dcolumn}% Align table columns on decimal point
\usepackage{bm}% bold math

\begin{document}
%\preprint{APS/123-QED}
\title{
Physically unclonable function using initial waveform of ring oscillators on 65 nm CMOS technology
}

\author{Tetsufumi Tanamoto}
 \email{tetsufumi.tanamoto@toshiba.co.jp}
\author{Satoshi Takaya}%
\affiliation{%
Corporate R \& D center, Toshiba Corporation,
Saiwai-ku, Kawasaki 212-8582, Japan
}%
\author{Nobuaki Sakamoto}%
\author{Hirotsugu Kasho}%
\affiliation{%
Toshiba Corporation Storage \& Electronic Devices Solutions Company, 
%580-1, Horikawa-cho, 
Saiwai-ku, Kawasaki 212-8520, Japan 
}%
\author{Shinichi Yasuda}
\author{Takao Marukame}
\author{Shinobu Fujita}
\author{Yuichiro Mitani}
\affiliation{%
Corporate R \& D center, Toshiba Corporation,
Saiwai-ku, Kawasaki 212-8582, Japan
}%

\date{\today}

\begin{abstract}
A silicon physically unclonable function (PUF) using ring oscillators (ROs) 
has the advantage of easy application in both an application specific integrated circuit (ASIC) 
and a field-programmable gate array (FPGA). 
Here, we provide a RO-PUF using the initial waveform of the ROs based on 65 nm CMOS technology.
Compared with the conventional RO-PUF, 
the number of ROs is greatly reduced and the time 
needed to generate an ID is within a couple of system clocks.
\end{abstract} 

\maketitle
\section{Introduction}
   In order to connect a huge number of devices as the Internet of Things (IoT)~\cite{IoT}, 
individual devices have to be manufactured at a low cost. 
This has also led to the technical issue of providing an inexpensive secure ID function 
in each device. Thus, research and development has focused on 
using the variability in each individual device as the chip fingerprint. 
A physically unclonable function (PUF) is considered to be one of the mechanisms 
of protecting personal information by providing a unique ID for each device at a low cost.
In the authentication process by a server, a PUF outputs a response ID to a challenge signal from the server.
If the PUF output is identified as the preregistered ID, the local device is authenticated.  
The origins of the PUF signal come from process variations of transistors and circuits. 
As transistor size has shrunk, various kinds of variation have emerged, 
resulting in an increasing number of proposals of many types of PUF\cite{arbiter,Lee,Guajardo,Holcomb,Kim,Marukame,Takaya,Takaya1,RRAM,Xie,Chen,Suh,Maiti0,Maiti1,Merli,Habib,Yin,Fischer,butterfly,FF-PUF,Yamamoto}. 
In particular, the PUF using ring oscillators (ROs)~\cite{Suh} has an advantage in that it can be implemented in commercial 
field-programmable gate arrays (FPGAs) without difficulty and its stability is proven~\cite{Maiti0,Merli,Yin}. 
However, previous RO-PUFs using the frequency difference between RO pairs [Fig.~1(a)] 
have disadvantages: the number of RO pairs corresponds to the length of the ID, 
with the result that a large number of ROs are required, and a continuous running of RO is undesirable 
from the viewpoint of power consumption. 
For example, Suh and Devadas~\cite{Suh} estimated that
128 pairs of oscillators (256 oscillators total) are required to generate 128 independent bits.
   In this study, we provide an RO-PUF using the waveform of ROs based on conventional 65 nm CMOS technology. The output waveform of ROs changes depending on each chip and provides
a long ID by a single pair of ROs.
Since the PUF ID is obtained after a couple of RO cycles, 
low power consumption is realized. 
Recently, in Ref.~\cite{tanaCASII} we have reported the implementation of a similar type of RO-PUF in commercial FPGA boards.
Because the placement and routing of circuits are automatically carried out, 
the circuit elements are arranged at a considerable distance from each other in FPGA. 
Thus, the PUF circuit cannot avoid inclusion of a noisy global wiring.
In ASIC devices, we can arrange the PUF circuit 
such that circuit elements are arranged in close proximity to each other 
to reduce unexpected noises. 
Here, we have designed the PUF circuits with a custom layout using standard logic cells, 
and found that the PUF performance is improved by ASIC chips. 
We also investigated the coupling effects of ROs in our PUF circuit.

The remainder of this paper is organized as follows: in Sect.~2, we describe
our proposal of an RO-PUF using initial waveforms of ring oscillators.
In Sect.~3, we present our experimental results of an RO-PUF based on 
65 nm CMOS chips.
In Sect.~4, we estimate the PUF performance quantitatively. 
Our conclusions are summarized in Sect.~5.

\begin{figure}
\begin{center}
\includegraphics[width=8.5cm]{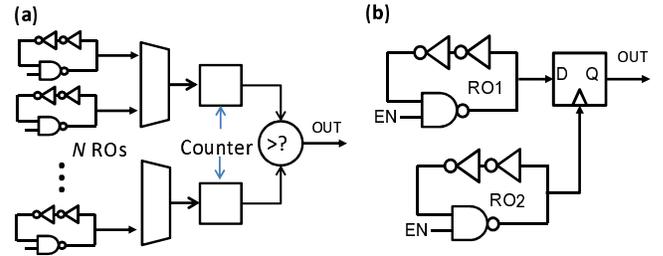}
\end{center}
\caption{(a) Conventional  ring oscillator (RO)-PUF in the literature. 
The frequencies of two ROs are compared by counters. 
(b) Proposed RO-PUF unit including two ROs. The output of RO1 is sampled by RO2. 
Because of the fast sampling by RO2, the output of (b) is also expected to be affected by the chip variations of FF.}
\label{f1}
\end{figure}

%%%%%%%%%%%%%%%%%%%%%%%%%%%%%%%%
\begin{figure}
\begin{center}
\includegraphics[width=8.5cm]{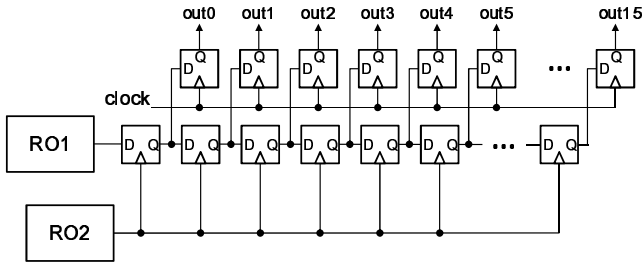}
\end{center}
\caption{Proposed PUF circuit including reading unit. 
The basic unit of Fig.~1(b) is connected to an FF array to read out the sampling by RO2. 
The number of FFs corresponds to the output ID length. This figure shows the case of a 16 bit output.}
\label{circuit}
\end{figure}

\section{PUF using waveform of ROs}
\subsection{Proposed PUF using waveform of ROs}
 The origin of a PUF is the process variation of transistors and circuits, 
and the previously proposed RO-PUFs show that process variations appear 
in the frequency of ROs~\cite{Suh}. 
The frequency difference among different ROs means that the period of one cycle 
between rising edges differs depending on each RO. 
Thus, it is natural to consider that the waveform difference of an RO can be used as the source of a PUF (wRO-PUF). 
The RO begins to oscillate when the supply voltage is applied or when the enable switch is turned on. 
When the initial output value of the RO is 0, 
the time to the first rising edge of oscillation signals differs depending on each RO. 
Figure~1(b) shows our proposed PUF unit consisting of two ROs. 
Because the frequency of RO1 ($\sim$1 GHz) is much higher than the clock frequency of the conventional 
circuit board ($\sim$50-100 MHz), in order to capture the oscillation signal of the RO1 at a higher time resolution, 
we use RO2 as the sampling clock of FF. 
Figure~\ref{circuit} shows the whole PUF circuit including reading FFs, 
where the number of FFs corresponds to the ID length. Here, experiments are carried out with 16 bits. 

\begin{figure}
\begin{center}
\includegraphics[width=8.5cm]{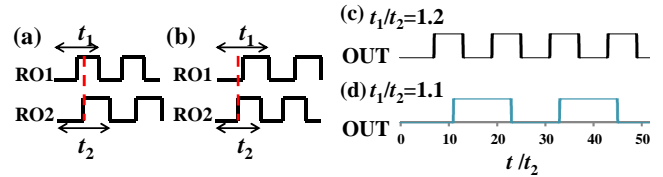}
\end{center}
\caption{(a), (b) Waveforms of outputs of RO1 and RO2. (c), (d) Output of Fig.~1(b).}
\label{waveform}
\end{figure}

The mechanism of our PUF is shown in Fig.~\ref{waveform}. 
When $EN=1$, the ROs start to oscillate from the initial value of 0. 
The time to the first rising edge differs depending on ROs. 
When the period of the waveform of RO1 is longer than that of RO2 ($t_1>t_2$), 
the initial value of the output in Fig.~\ref{waveform}(a) is 0. 
When $t_1<t_2$, the initial value of the output in Fig.~\ref{waveform}(a) is 1. 
This mechanism resembles that of the conventional RO-PUFs. 
Moreover, because the waveforms are analog data, 
the relative difference $|t_1-t_2|/ t_2$ between two waveforms is the origin of new degrees of freedom 
when they are digitized. Figures~\ref{waveform}(c) and \ref{waveform}(d) show examples where the output signal (OUT) pattern 
of the wRO-PUF changes depending on the relative change $t_1/t_2$ even when we limit $t_1>t_2$. 
$t_1/t_2=1.1$ and $t_1/t_2=1.2$ for the difference, even when we limit $t_1>t_2$.
   
\subsection{Effect of coupling between two ROs}
 In the basic unit of our PUF [Fig.~1(b)], the relative frequency between two ROs determines the waveform of the PUF-ID. 
As Ref.~\cite{Maneatis} shows that the coupling of ROs improves the delay control, 
we have also implemented modified PUF circuits whose units are shown in Fig.~\ref{couple}. 
Figure~\ref{couple}(a) shows a PUF circuit of electrically coupled ROs, 
and Fig.~\ref{couple}(b) shows that of capacitively coupled ROs. 

%%%%%%%%%%%%%%%%%%%%%%%%%%%%%%%%

\section{PUF chip using 65 nm CMOS process}
\subsection{Hamming distance evaluation of RO-PUF using waveform}
Figure~\ref{chip} shows a photomicrograph of an RO-PUF chip obtained using 65 nm CMOS technology where the macrosize is 273$\times$28 $\mu$m${}^2$. 
We measured 10 chips. ID is determined by the pattern of the wRO-PUF output that appears most frequently. 
The most important metric for estimating PUF performance is the Hamming distance (HD), 
which is the number of different bits between two outputs. 
The intra-HD represents the reproducibility of each chip, and the 
inter-HD represents the uniqueness between different chips.
A sharp distribution at around HD=0 and a sufficient gap between the intra-HD 
and the inter-HD indicate the ideal HD distribution. 
The ideal inter-HD is distributed around half of the ID length. 

\begin{figure}
\begin{center}
\includegraphics[width=8.5cm]{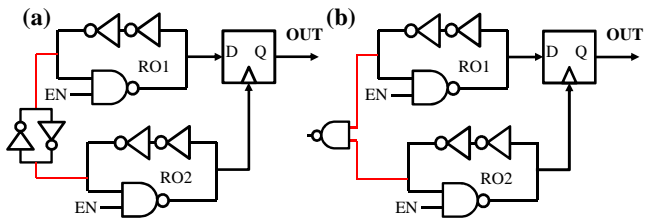}
\end{center}
\caption{Coupling of two ROs. (a) Coupling by the inverter loop. (b) 
Capacitive coupling by NAND cell. It is found that the coupling in (a) is strong and it does not work as a PUF. 
We can see the effect of the coupling in (b). }
\label{couple}
\end{figure}

\begin{figure}
\begin{center}
\includegraphics[width=8cm]{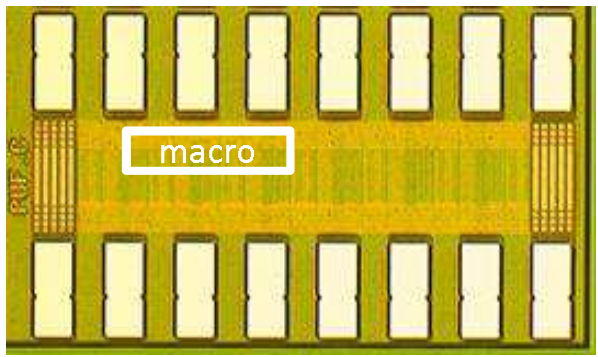}
\end{center}
\caption{Photomicrograph of the chip obtained using 65~nm CMOS technology.}
\label{chip}
\end{figure}

\begin{figure}
\begin{center}
\includegraphics[width=8cm]{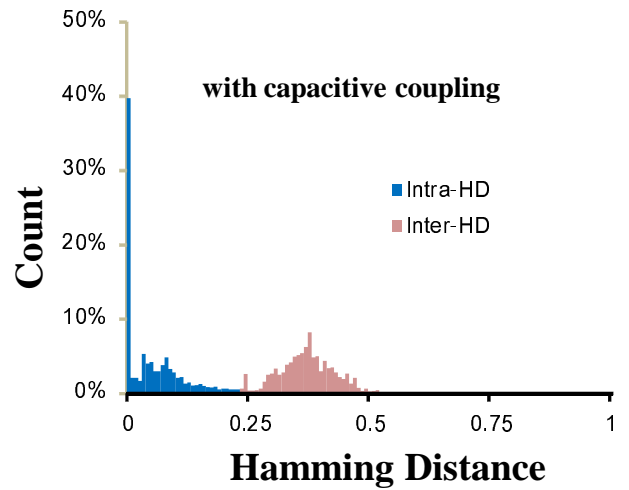}
\end{center}
\caption{Hamming distances of the RO-PUF using the initial waveforms in Fig.~\ref{couple}(b). 
 Ten chips are measured where 5000 ID output data are estimated.
Eight RO pairs of a 32 bit ID are combined. 
The (31, 16, 7) binary Bose-Chaudhuri-Hocquenghem (BCH) error-correcting code is applied.}
\label{HDw}
\end{figure}

First, let us show the result of the modified PUF in Fig.~\ref{couple}.  
It is found that the RO-PUF formed by cross-coupled inverters [Fig.~\ref{couple}(a)] only output 
fixed values and did not work as a PUF. This shows that this coupling is very strong for detecting the variation of the PUF output. 
Figure~\ref{HDw} shows the measured results for the modified PUF circuit in Fig.~\ref{couple}(b). 
Here, samplings are carried out by repeating $EN=0$ and $EN=1$.  
Figure~\ref{HDw} shows that the intra-HD of the wRO-PUF with a capacitive coupling is not ideal because 
the result indicated that the same ID is not always produced with a high probability. 

\begin{figure}
\begin{center}
\includegraphics[width=8cm]{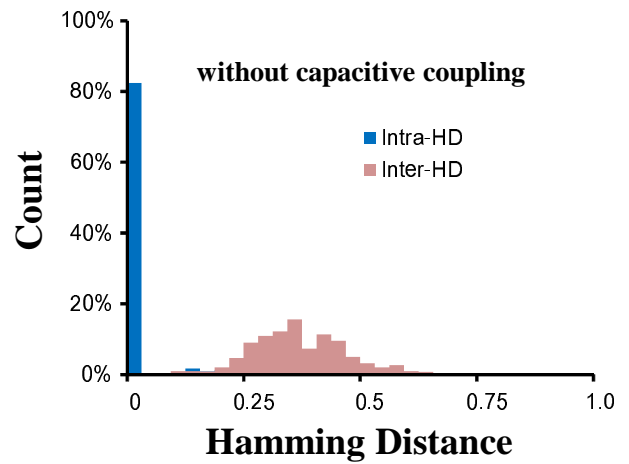}
\end{center}
\caption{Hamming distances of the RO-PUF using the initial waveforms in Fig.~\ref{circuit}. 
 Ten chips are measured where 5000 ID output data are estimated.
Two RO pairs of a 32 bit ID are combined.
The (31, 16, 7) binary BCH error-correcting code is applied.}
\label{HDwo}
\end{figure}

Figure~\ref{HDwo} shows the result of the HD in Fig.~\ref{circuit}.  
We can see that the identification of each chip is excellent (HD=0) 
and well distinct from those of the other chips. 
These results show that the simplest form of the wRO-PUF is the best form of the wRO-PUF.

%20170122
%%%%%%%%%%%%%%%%%%%%%%%%%%%%%%%%%%%%%%%%%%%%%%%%%

%%%%%%%%%%%%%%%%%%%%%%%%%%%%%%%%%%%%%%%%%%%%%%%%%

\subsection{Voltage dependence}
\begin{figure}
\begin{center}
\includegraphics[width=8cm]{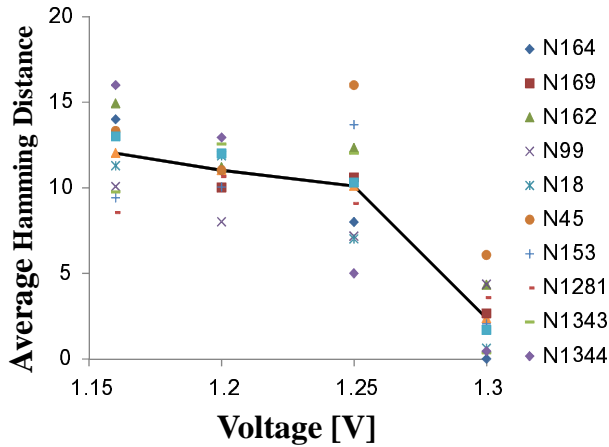}
\end{center}
\caption{Voltage dependence of the average Hamming distance from 1.3 V operation.
Ten chips are evaluated. }
\label{voltage}
\end{figure}

\begin{figure}
\begin{center}
\includegraphics[width=8cm]{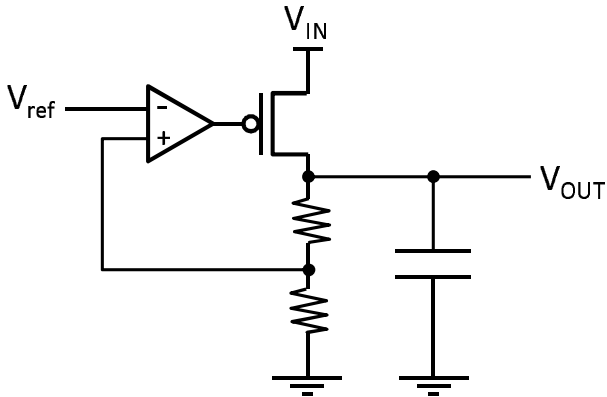}
\end{center}
\caption{Typical voltage regulator to stabilize supply voltage. }
\label{regulator}
\end{figure}

As shown in Ref.~\cite{Sayed}, RO frequency changes as the physical environment changes.
Figure~\ref{voltage} shows the shift of the average Hamming distance estimated from a 1.3 V operation over ten chips of the wRO-PUF without coupling.
We can see a linear dependence between the voltage difference and the shift in the average Hamming distance.
This is the same tendency as the temperature dependence shown in Ref.~\cite{Sayed}.
Because the ID should be unique, a change of the ID is undesirable. 
%%%%%%%%%%%% new
%{\bf 
To solve this problem, 
we can utilize the linear dependence of the voltage during the ID shift by monitoring voltage.
That is, the ID is corrected depending on the difference in the measured voltage from the reference voltage.
%Otherwise, a feedback circuit to control the PUF output will be required.
Otherwise, a feedback circuit for stabilizing the supply voltage will be required~\cite{Miliken}.
Figure~\ref{regulator} shows a schematic of a typical voltage regulator.
Because the change in frequency depending on the change in voltage is  
a general aspect of an RO-PUF, these kinds of additional circuits will be required 
in other conventional RO-PUF circuits.
%}
In any event, this is a problem to be addressed in future studies.  

%%%%%%%%%%%%%%%%%%%%%%% Fig Voltage dependence
 
%%%%%%%%%%%%%%%%%%%%%%% Fig Voltage regulator

%%%%%%%%%%%%%%%%%%%%%%%%%%%% new
%{\bf
\subsection{Comparison with previous RO-PUFs}
We would like to quantitatively compare our RO-PUF with the previously proposed RO-PUF~\cite{Suh}.
First, let us compare the areas of both PUFs, when 128 bit outputs are required.
For our wRO-PUF circuit, we prepare two sets of RO pairs [Fig.~\ref{circuit}], each of which outputs more than 64 bits.
Then, our wRO-PUF has 4 ROs and 64 (=$16 \times 2 \times 2$) FFs.
If each RO has 20 transistors, and 30 transistors are used for each FF,
about 2000 transistors are present in our wRO-PUF circuit.
If we reduce the number of FFs, which dominate the area of our wRO-PUF, we can further
reduce the area of our circuit. 
Suh and Devadas~\cite{Suh} showed that 35 ROs are used to obtain 133 bits.
Their typical RO-PUF has two MUXs, two counters, and a comparison circuit.
When we assume that each MUX consists of combined 4-input MUXs, each of which has about 30 transistors, 
21 (=16+4+1) 4-input MUXs are required, resulting in 1260 transistors for the MUX parts.
If 16 bit counters are used, 32 FFs with additional circuits are required, resulting in 
960 transistors being present in this part.
When full adders are used to compare two RO outputs, and each full adder has 20 transistors, there are 320 transistors.  
Thus, the previously proposed PUF has about 3240 transistors  (Table I).  
 
Next, let us compare power consumptions.
For the previous PUF in Ref.~\cite{Suh}, 16 bit counters are operated to obtain a bit.
Therefore, the previous RO-PUF has to run an RO pair for at least 2048 (=$16\times128$) cycles of the system clock.
In contrast, an RO pair of our wRO-PUF acquires 16 bits during a cycle of the system clock. 
Thus, the RO pair should run only 8 (=128/16) clocks in our wRO-PUF, resulting in a great reduction in power (Table I).
%}

%{\bf   %new 20161219
%%%%%%%%%%%%%%%%%%%%%%%%%%%%%%%%%%%%%%%%%%%%%%%%%
\begin{table}
\begin{center}
Table I. 
{Comparison of area (number of transistors) and clock cycle (128 bit output) }. 
\begin{tabular}{|c|c|c|}\hline
%%% 1st line  ================* 
&   Area  %(45nm)
&   Clock cycle
\\ \hline
%%% 2nd line  ================*
%1-col
Our wRO-PUF 
& 2000 
& 8
\\
%%% 3rd line  ================*
Previous RO-PUF
& 3240  
& 2048 
%%% 4th line  ================*
%&\textcolor{red}{ 50} \%
\\ \hline
\end{tabular}
\end{center}
\end{table}
%}

\section{PUF performance evaluation}
We evaluate the PUF performance of our wRO-PUF on the basis 
of the metrics discussed in Refs.~\cite{Maiti3} and \cite{Hori}.
The results of the evaluation with and without coupling are shown in Table II.
Here, uniqueness is calculated using
\begin{equation}
{\rm Uniqueness}=\frac{2}{N(N-1)}\sum_{i=1}^{N-1}\sum_{j=i+1}^N
\frac{HD(R_i,R_j)}{L}\times 100\%,
\end{equation}
where $N$ is the number of tested chips, $L$ is  
the length of the ID, and $R_i$ is the ID pattern that appears most frequently.
The reliability of the $i$-th chip is calculated using
\begin{equation}
{\rm Reliability}_i= \left[1-\frac{1}{T}\sum_{t=1}^{T}\frac{HD(R_i,R'_{i,t})}{L} \right]\times 100\%,
\end{equation}
where $T$ is the number of times that sampling is done and $R'_{i,t}$ is the $t$-th output.  
Uniformity is calculated using 
\begin{equation}
{\rm Uniformity}_{i,n}=\frac{1}{L}\sum_{l=1}^{L} b_{i,n,l} \times 100 \%,
\end{equation}
%{\bf 
where $b_{i,n,l}$ is the $l$-th binary bit of an $n$-th response from a chip $i$. %}
We average Uniformity${}_{i,n}$ over many responses.
The average results over all samples are listed in Table II.

It seems that the results shown in Table II are inferior to those shown in Refs.~\cite{Maiti3} and \cite{Hori}.
This is considered to be because the number of ROs is much smaller 
than those in the literature. 
Here, the number of RO pairs in the coupling case is four and that in the case without coupling is one.
In general, as the number of ROs increases, these metrics are considered to improve,
because the complexity increases as the number of ROs increases.
%{\bf 
The PUF performance of an RO pair in the present ASIC chip seems to be better than 
that of FPGA. Therefore, two or three RO pairs will improve the PUF performance. % }
%In other words, as the number of ROs increases, the bare characteristics of RO-PUF 
%become obscured. 
Thus, from these results, we can see the original properties of our RO-PUF.
The effect of coupling can be seen in the uniqueness estimation. 
The uniqueness of our wRO-PUF without coupling is close to the ideal value, resulting in the 
excellent performance of our wRO-PUF without coupling.
In any event, the improvement in PUF performance is a future problem.

%%%%%%%%%%%%%%%%%%%%%%%%%%%%%%%%%%%%%%%%%%%%%%%%%
\begin{table}
\begin{center}
Table II. 
{PUF performance measured on the two types of chip (in \%)}. 
\begin{tabular}{|c|c|c|c|}\hline
%%% 1st line  ================* 
&  With coupling %(45nm)
&  Without coupling %(90nm)
&  Ideal value
\\ \hline
%%% 2nd line  ================*
%1-col
Uniformity 
& 40.82 %C:\Users\e1093010\Documents\PUF\TEG\20160525\uniformity_grep.xlsx
& 38.75 %C:\Users\e1093010\Documents\PUF\PUF2015csrd\20160912\uniq1.3v.xlsx
&\ 50 
\\
%%% 3rd line  ================*
Reliability
& 94.67  %C:\Users\e1093010\Documents\PUF\TEG\20160525HD_resultReliabilityECC_1605251.xlsx
& 94.48 %C:\Users\e1093010\Documents\PUF\PUF2015csrd\20160912\HD_resultHDaaECC1.2vON
& 100 
\\
%%% 4th line  ================*
%1 col
Uniqueness
& 37.22  %C:\Users\e1093010\Documents\PUF\TEG\20160525\uniqECC_160525.xlsx
& 48.81  %C:\Users\e1093010\Documents\PUF\PUF2015csrd\20160912\uniq1.3v.xlsx
& \ 50 
%&\textcolor{red}{ 50} \%
\\ \hline
\end{tabular}
\end{center}
\end{table}
%w/o 8はデバイス数１６はビット数 Z=3
%w/ 15はデバイス数１６はビット数

%%%%%%%%%%%%%%%%%%%%%%%%%%%%%%%%%%%%%%%%%%%%%%%%% 
\section{Conclusions}
   We proposed an RO-PUF using the initial waveform (wRO-PUF) based on a 65 nm CMOS circuit. 
The initial waveform of an RO is effectively sampled by other ROs. 
No coupling between two ROs is required. The simple circuit [Figs.~1(b) and 2] is proved to be best for the PUF-ID. 
Compared with the conventional RO-PUF using a frequency difference, a small number of ROs 
are used with a couple of system clock operations, resulting in a low power consumption and a small area.

\acknowledgments
T.T. thanks A. Nishiyama, M. Koyama, K. Muraoka, S. Shimizu, 
Y. Komano, H. Noguchi, T. Kanesige, T. Ootuki,
and T. Yamakawa for helpful discussions.


\begin{thebibliography}{99}
\bibitem{IoT}
X. Liu, J. Zhou, C. Wang, K. Chang, J. Lan, L. Liao, Y. Lam, Y. Yang, B. Wang, X. Zhang, W. Goh, T. Kim, and M. Je, 
%"An Ultra-Low-Voltage Sensor Node Processor with Diverse Hardware Acceleration 
%and Cognitive Sampling for Intelligent Sensing Applications," 
IEEE Trans. Circuits Syst. II, {\bf 62}, 1149 (2015).
%no.12, Dec. 2015, pp. 1149-1153.   

%1st PUF
%\bibitem{Gassend}
%B. Gassend, D. Clarke, M. van Dijk, and S. Devadas, "Controlled
%Physical Random Functions",
%18th Annual Computer Security Applications Conference (IEEE
%Computer Society, Washington, DC, 2002), p. 149

%Arbiter
\bibitem{arbiter} %ok
D. Lim, J. Lee, B. Gassend, G. E. Suh, M. van Dijk, and S. Devadas, 
%D. Lim, {\it et al.}, %J. Lee, B. Gassend, G. Suh, M. van Dijk, and S. Devadas, 
%``Extracting secret keys from integrated circuits,” {\it I
IEEE Trans.VLSI Syst. {\bf 13}, 1200 (2005).
%no. 10, Oct. 2005, pp. 1200-1205.

\bibitem{Lee} %o
J. W. Lee, D. Lim, B. Gassend, G. E. Suh, M. van Dijk, and S. Devadas, 
%``A Technique to Build a Secret Key in Integrated
%Circuits for Identification and Authentication Application", in {\it 
IEEE Symp. VLSI Circuits, 2004, p. 176.
%, pp. 176-159.

%SRAM
\bibitem{Guajardo} %ok
J. Guajardo, S. S. Kumar, G.-J. Schrijen, and P. Tuyls, 
%“FPGA intrinsic PUFs and their use for IP protection”, in {\it 
9th Int. Workshop Cryptographic Hardware and Embedded Systems (CHES'07), 2007, p. 63.

\bibitem{Holcomb}
D. E. Holcomb, W. P. Burleson, and K. Fu, 
%``Power-up SRAM State as an Identifying Fingerprint and Source of True Random Numbers”, {\it 
IEEE Trans. Comput. {\bf 58}, 1198 (2009). 

%--memory puf
\bibitem{Kim}
M. S. Kim, D. I. Moon, S. K. Yoo, and S. H. Lee, 
%``Investigation of Physically Unclonable Functions Using Flash Memory for Integrated Circuit", 
IEEE Trans. Nanotechnol. {\bf 14}, 384 (2015).

\bibitem{Marukame}
T. Marukame, T. Tanamoto, and Y. Mitani,
%``Extracting Physically Unclonable Function From Spin Transfer Switching Characteristics in Magnetic Tunnel Junctions", {\it 
IEEE Trans. Magn. {\bf 50},  1 (2014). 

\bibitem{Takaya}
S. Takaya, T. Tanamoto,  H. Noguchi, K. Ikegami, K. Abe, and S. Fujita,
Ext. Abstr. Int. Conf. Solid State Devices and Materials (SSDM2016), 2016, B-2-05.

\bibitem{Takaya1}
S. Takaya, T. Tanamoto,  H. Noguchi, K. Ikegami, K. Abe, and S. Fujita,
to be published in Jpn. J. Appl. Phys. (SS16089)

\bibitem{RRAM}
A. Chen,
%``Utilizing the Variability of Resistive Random Access Memory to Implement Reconfigurable Physical Unclonable Function", 
IEEE Electron Device Lett. {\bf 36}, 138 (2015).

\bibitem{Xie}
Y. Xie, X. Xue, J. Yang, Y. Lin, Q. Zou, R. Huang, and J. Wu,
%“A Logic Resistive Memory Chip for Embedded Key Storage With Physical Security”, 
IEEE Trans. Circuits Syst. II, {\bf 63}, 336 (2015).

\bibitem{Chen}
J. Chen, T. Tanamoto, H. Noguchi, and Y. Mitani,
%``Further investigations on traps stabilities in random telegraph signal noise and the application to a novel concept physical unclonable function (PUF) with robust reliabilities", 
IEEE Symp. VLSI Technology, 2015, p. 40.

%RO
\bibitem{Suh} %ok
G. E. Suh and S. Devadas, 
%“Physical unclonable functions for device authentication and secret key generation”, in {\it 
44th Ann. Design Automation Conf. (DAC'07), 2007, p. 9.

\bibitem{Maiti0} %ok
A. Maiti, J. Casarona, L. McHale, and P. Schaumont, 
%“A large scale characterization of RO-PUF”, in {\it 
IEEE Symp. Hardware-Oriented Security and Trust (HOST), 2010, p. 94.
%}, 2010, pp. 94-99.

\bibitem{Maiti1} %ok
A. Maiti and P. Schaumont, 
%``Improving the quality of
%a physical unclonable function using configurable ring oscillators", In {\it 
19th Int. Conf. Field Programmable Logic and Applications 
(FPL'09), 2009, p. 703.

%\bibitem{Gehrer}
%Reconfigurable PUFs for FPGA-based SoCs
%Stefan Gehrer Robert Bosch GmbH

\bibitem{Merli} %o
D. Merli, F. Stumpf, and C. Eckert, 
%“Improving the quality of ring oscillator PUFs on FPGAs”, in {\it 
5th Workshop Embedded Systems Security (WESS' 10), 2010.

\bibitem{Habib} %o
B. Habib, K. Gaj, and J.P. Kaps, 
%“FPGA PUF Based on Programmable LUT Delays”, in {\it Proc. 
Euromicro Conf. Digital System Design (DSD'13), 2013, p. 697.

\bibitem{Yin}
C.-E. Yin and G. Qu, %“Lisa: Maximizing RO PUF’s secret extraction,”
%in {\it 
IEEE Symp. Hardware-Oriented Security and Trust (HOST), 2010, p. 100.
%}, 2010, pp. 100-105.
%}
\bibitem{Fischer}
L. Bossuet, X. T. Ngo, Z. Cherif, and V. Fischer, 
%"A PUF Based on a Transient Effect Ring 
%Oscillator and Insensitive to Locking Phenomenon,". {\it 
IEEE Trans. Emerging Top. Comput. {\bf 2}, 30 (2014).


\bibitem{butterfly}  %ok
S. Kumar, J. Guajardo, R. Maes, G.-J. Schrijen, and P. Tuyls, 
%“Extended abstract: The butterfly puf protecting IP on every FPGA”, in
%{\it Proc. IEEE HOST
IEEE Symp. Hardware-Oriented Security and Trust (HOST 2008), 2008, p. 67.

\bibitem{FF-PUF}
R. Maes, P. Tuyls, and I. Verbauwhede, 
%``Intrinsic PUFs from flip-flops on reconfigurable devices", in {\it 
Workshop Information and System Security (WISSec), 2008, p. 17. 
%, pp. 17-26.

\bibitem{Yamamoto} %o
D. Yamamoto, K. Sakiyama, M. Iwamoto, K. Ohta, T. Ochiai, M. Takenaka, and K. Itoh, 
%``Uniqueness enhancement of puf responses based on the locations of random
%outputting RS latches", in {\it Proc. IEEE HOST
13th Int. Workshop Cryptographic Hardware and Embedded Systems (CHES'11), 2011, p390.
%}, 2011, pp. 390-406.

%\bibitem{Liu}
%\textcolor{red}{
%Liu Dongsheng, Liu Zilong, Li Lun, and Zou Xuecheng, 
%D. Liu {\it et al}., %Dongsheng, Liu Zilong, Li Lun, and Zou Xuecheng, 
%"Low-Cost Low-Power Ring Oscillator-based Truly Random Number Generator for Encryption on Smart Cards", 
%{\it IEEE Trans. Circuits Syst. II, Exp. Briefs}, Vol. 63, June, 2016,pp. 608 - 612.
%}

%\bibitem{Robson}
%\textcolor{red}{
%S. Robson, {\it et al}., % Bosco Leung ; Guang Gong
%"Truly Random Number Generator Based on a Ring Oscillator Utilizing Last Passage Time"
%{\it IEEE Trans. Circuits Syst. II, Exp. Briefs,} Vol.61, Dec. 2014,
%pp. 937 - 941.
%}


\bibitem{tanaCASII}
T. Tanamoto, S. Yasuda, S. Takaya, and S. Fujita,
%"Physically Unclonable Function using Initial
%Waveform of Ring Oscillators",
to be published in IEEE Trans. Circuits Syst. II.

\bibitem{Maneatis}
J. G. Maneatis and M.A. Horowitz,
%"Precise delay generation using coupled oscillators"
IEEE J. Solid-State Circuits {\bf 28}, 1273 (1993).

\bibitem{Sayed}
M. A. Sayed, and P. H. Jones, 
%“Characterizing non-Ideal Impacts of Reconfigurable Hardware Workloads on Ring Oscillator-based Thermometers”, 
%in {\it ReConFig 2011 
IEEE Int. Conf. Reconfigurable Computing and FPGAs (ReConFig2011), 2011, p. 92.
%}, 2011, pp. 92-98.

\bibitem{Miliken}
R. J. Milliken, J. Silva-Martínez, and E. Sánchez-Sinencio, 
%Full On-Chip CMOS Low-Dropout Voltage Regulator
IEEE Trans. Circuits Syst. I {\bf 54}, 1879 (2007). 

\bibitem{Maiti3}
%Abhranil Maiti, Member, IEEE, and Patrick Schaumont, Senior
A. Maiti and P. Schaumont, 
%``The Impact of Aging on a Physical Unclonable Function",
IEEE Trans. VLSI Syst., {\bf 22}, 1854 (2013).
%, pp.1854-1864.

\bibitem{Hori} %ok
Y. Hori, T. Yoshida, T. Katashita, and A. Satoh, 
%“Quantitative and statistical performance evaluation of arbiter physical unclonable
%functions on FPGAs”, in 
IEEE Int. Conf. Reconfigurable Computing and FPGAs (ReConFig2010), 2010, p. 298.
%}, 2010, pp. 298-303.

%\bibitem{Yang}
%Kaiyuan Yang, Qing Dong, David Blaauw, Dennis Sylvester
%K. Yang {\it et al}., %Q. Dong, D. Blaauw, and D. Sylvester,
%``A Physically Unclonable Function with BER$<10^{-8}$ for
%Robust Chip Authentication Using Oscillator Collapsein 40nm CMOS",
%{\it IEEE ISSCC},  2015, pp. 254-255.

\end{thebibliography}
\end{document}